\documentclass[10pt, conference]{IEEEtran}
\IEEEoverridecommandlockouts
\usepackage{cite}
\usepackage{amsmath,amssymb,amsfonts}
\usepackage{algorithmic}
\usepackage{graphicx}
\usepackage{textcomp}
\usepackage{xcolor}
\newcommand*{\Resize}[2]{\resizebox{#1}{!}{$#2$}}%

\def\BibTeX{{\rm B\kern-.05em{\sc i\kern-.025em b}\kern-.08em
    T\kern-.1667em\lower.7ex\hbox{E}\kern-.125emX}}
\begin{document}

\title{Unsupervised Generative Adversarial Alignment Representations for Sheet music, Audio and Lyrics\\
}
\author{Donghuo Zeng,\quad Yi Yu,\quad Keizo Oyama \\
National Institute of Informatics, SOKENDAI \\
Tokyo, Japan \\
\{zengdonghuo, yiyu, oyama\}@nii.ac.jp}

\maketitle

\begin{abstract}
Sheet music, audio, and lyrics are three main modalities during writing a song. In this paper, we propose an unsupervised generative adversarial alignment representation (UGAAR) model to learn deep discriminative representations shared across three major musical modalities: sheet music, lyrics, and audio, where a deep neural network based architecture on three branches is jointly trained. 
In particular, the proposed model can transfer the strong relationship between audio and sheet music  to audio-lyrics and sheet-lyrics pairs by learning the correlation in the latent shared subspace. 
We apply CCA components of audio and sheet music to establish new ground truth. The generative (G) model learns the correlation of two couples of transferred pairs to generate new audio-sheet pair for a fixed lyrics to challenge the discriminative (D) model. The discriminative model aims at distinguishing the input which is from the generative model or the ground truth. The two models simultaneously train in an adversarial way to enhance the ability of deep alignment representation learning.  Our experimental results demonstrate the feasibility of our proposed UGAAR for alignment representation learning among sheet music, audio, and lyrics.
\end{abstract}

\begin{IEEEkeywords}
Adversarial learning, representation learning, cross-modal retrieval
\end{IEEEkeywords}

%
\IEEEpeerreviewmaketitle
\section{Introduction}
With the rapid growth of music contents including users' annotations emerging in Internet, it is becoming very important to learn common semantics of music alignment representation for facilitating cross-modal music information retrieval. For example, when we input "kids" as query to search a song’s audio, video or lyrics, what we expected is an audio that exists kids' voice, video contains kids or lyrics has semantic kids' information. Such semantic concept in audio, video and lyrics are based on explicit concept "kids", which is defined by users. In this paper, sheet music, audio and lyrics are implicitly aligned by high-level semantic concepts, so we develop a content-based representation learning approach by learning alignment across these modalities for retrieval task. The approach ensures the search engine to find the exactly paired music data, without involving the problems of deviation of users' preference.

The main challenge of representation learning across different musical modalities is the heterogeneous gap. In previous works, representation learning for musical cross-modal retrieval focus more on two modalities to bridge the modality gap, such as audio-sheet music~\cite{ThomasFMC12}, which achieved success in musical cross-modal retrieval domain. A classical method series is the CCA-based approaches, which aims at finding transformation to optimize the correlation between the input pairs from two different variable sets. In order to be beneficial from CCA and rank loss for two modalities aligned representation learning, CCA layer~\cite{dorfer2018end} combines the existing representations learning like pairwise loss,  with the optimal projections of CCA to learn representation between the short snippets of music and the corresponding part of sheet music for the content-based sheet music-audio retrieval scenarios.

\begin{figure*}
    \centering
    \includegraphics[width=15cm, height=7.6cm]{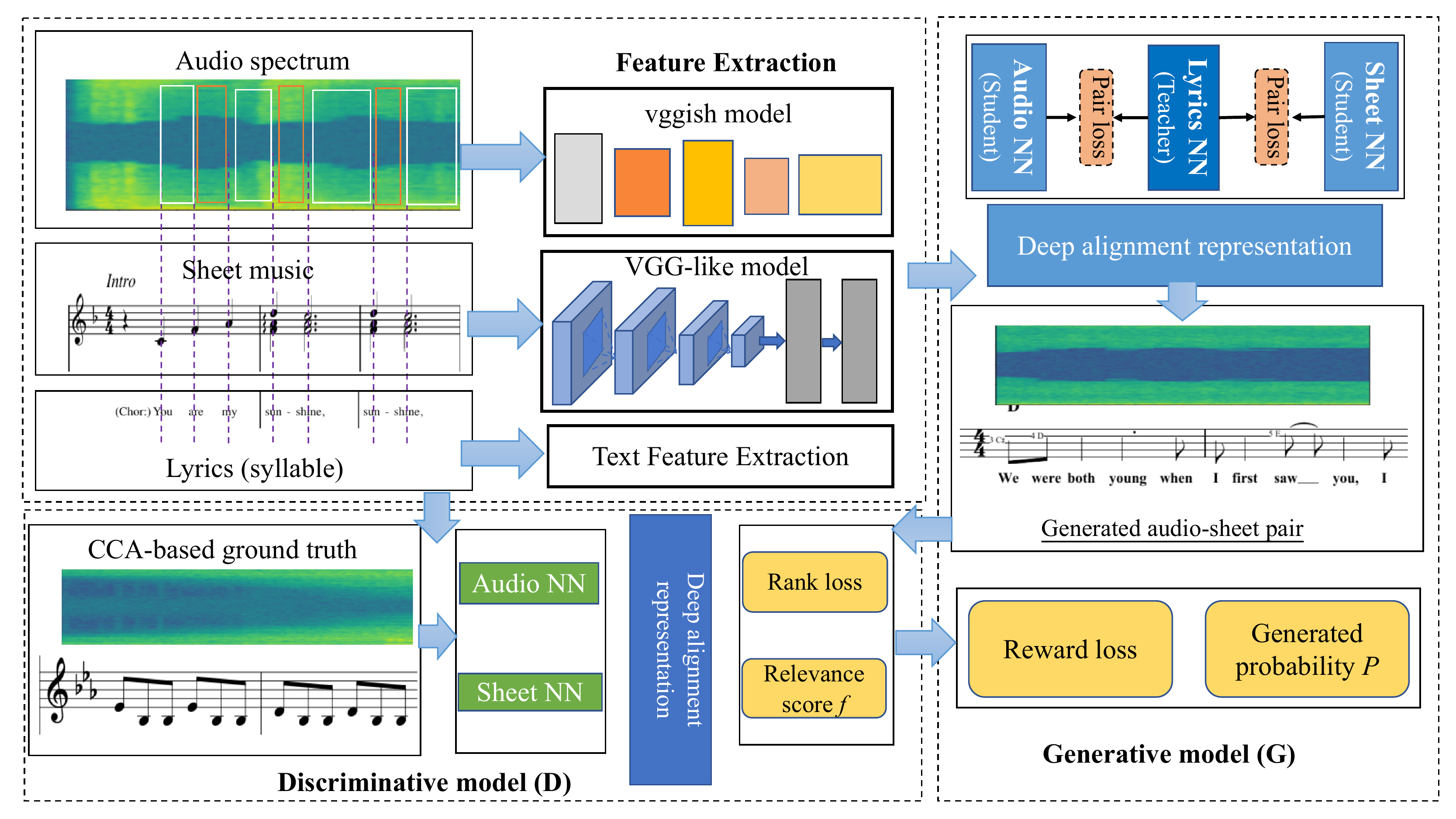}
    \caption{Our proposed architecture, which includes three parts: feature extraction, Generative model and Discriminative model.}
    \label{fig:model}
\end{figure*}
Learning aligned representations between two modalities has progressively been arranged in cross-modal retrieval~\cite{yu2018category, yu2019deep, zeng2018audio}, such as learning temporal relation~\cite{FujiharaG12} between audio and lyrics for various applications, deep sequential correlation~\cite{yu2019deep} between audio and lyrics for cross-modal retrieval. However, it is hard to satisfy the requirement of real multimodal information retrieval when retrieving one modality by the other two modalities. The goal of this paper is to learn a robust alignment representation for sheet music, audio and lyrics by unsupervised learning, and explore the representations for three groups of cross-modal retrieval tasks to evaluate the performance of our architecture.

Little research has been conducted on the content-based alignment representation learning among musical modalities: audio, sheet music and lyrics, due to the limited available musical dataset. In this paper, we collected a musical dataset with three modalities, including musical audio, sheet music and lyrics. In the dataset, audio and sheet music are paired because they are generated by music notes. This paper has achieved two main contributions: i) our architecture can transfer the audio-sheet music pair to audio-lyrics and sheet music-lyrics pairs by generative adversarial networks (GANs), and some results achieved by our approach on the Musical Ternary Modalitie (MTM) dataset prove the feasibility of learning aligned across three modalities by transferring one close relationship to the other two couples of relationships. ii) we combine the objective of existing CCA projection with the optimal representations of GANs. In detail, we establish a new ground truth based on the CCA embedding and explore generative model to generate new audio-sheet music pair, the discriminative model tries to distinguish the input is from G model or ground truth. During the adversarial learning, the G model can generate more discriminative and aligned representation for lyrics, audio and sheet music.

\section{PROPOSED METHOD}

\subsection{Generative Model}
We take advantage of the close relationship of audio and sheet music. The generative model is to generate a new audio and sheet music pair with a fixed lyrics to fool the discriminative model. 

\subsubsection{Alignment by Model Transfer}
Aspired by Teacher-Student model~\cite{AytarVT17}, we assume audio and sheet music are Student models, lyrics is the Teacher model. Our model tries to establish new aligned representations for all of them. Let $x_{i}$ be a data point from audio set $x$, the corresponding data point $y_{i}$ and $z_{i}$ are from sheet music set $y$ and lyrics set $z$.
The new generated aligned representations $f(x_{i})$ and $g(y_{i})$ of audio and sheet music from our model are trained with lyrics $z_{i}$.
Because the three modalities are synchronized, we can learn $h(z_{i})$ model for lyrics $z_{i}$ to predict the representation of $f(x_{i})$ and $g(y_{i})$. The Teacher-Student model is achieved by the KL-divergence~\cite{AytarVT17} loss:
\begin{eqnarray}
  \sum_{i}^{n}D_{KL}(h(z_{i})||f(x_{i}))=\sum_{j}^{n}h(z_{i})log\frac{h(z_{i})}{f(x_{i})} \\
  \sum_{i}^{n}D_{KL}(h(z_{i})||g(y_{i}))=\sum_{i}^{n}h(z_{i})log\frac{h(z_{i})}{g(y_{i})}
\end{eqnarray}

The model transfer enhances the audio and sheet music to learn the discriminative representation as lyrics. To reinforce three components of a ternary have similar representations, we enable an alignment across different modalities by generative probability.
\begin{figure*}
    \centering
    \includegraphics[width=14.5cm, height=5.2cm]{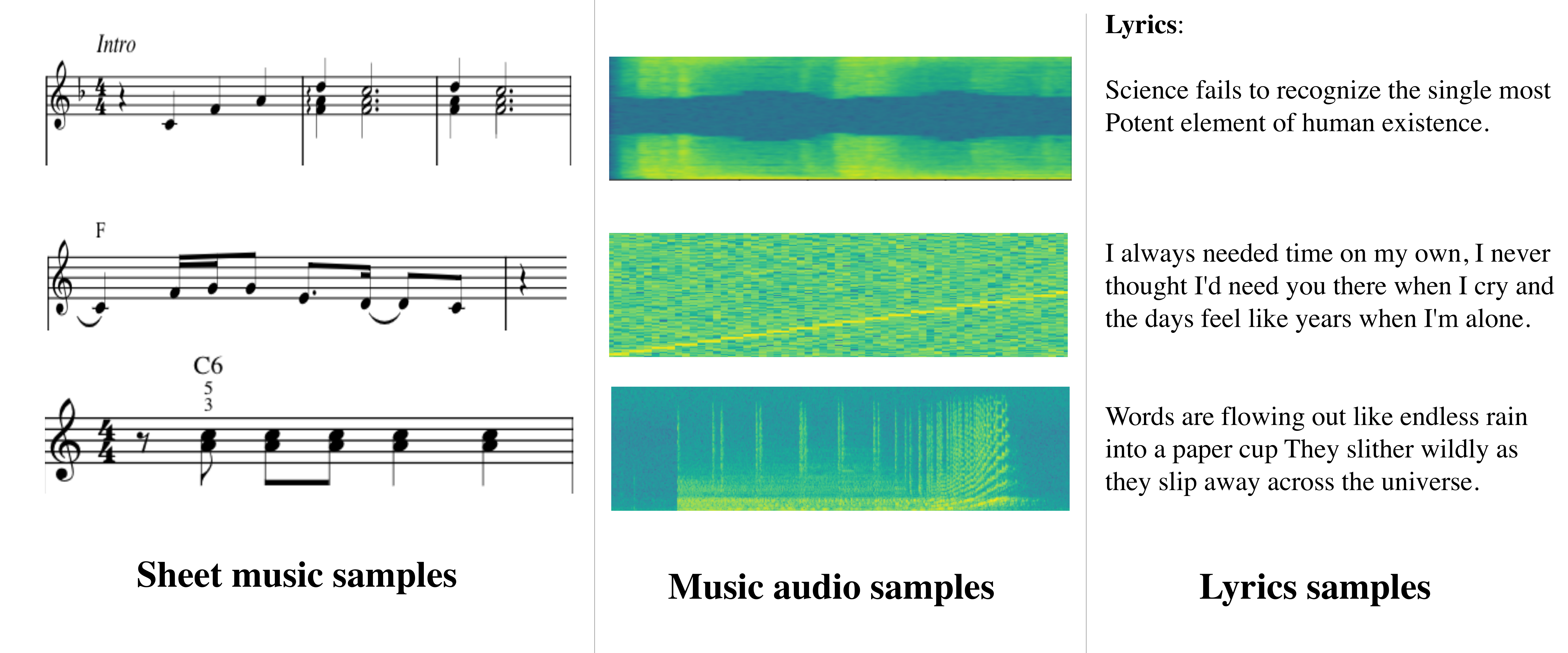}
    \caption{Samples of sheet music, audio, and lyrics in our dataset}
    \label{fig:samples}
\end{figure*}
\subsubsection{Alignment by Generative Probability}
given a lyrics, generative model G aims at fitting the contribution over the lyrics-audio and lyrics-sheet music pairs in a common space by mapping function $F(x_{i})$, $G(y_{i})$ and $H(z_{i})$ for audio $x_{i}$, sheet music $y_{i}$ and lyrics $z_{i}$. Then, the pairs of informative audio and sheet music are selected to test the ability of discriminative model D. The generative probability of G is $p_{\theta}(x, y|z)$ is the foundation of selecting relevant audio-sheet music pair from unpaired data with lyrics. For instance, given a lyrics query $z_{i}$, the generative model tries to select relevant audio $x_{j}$ from $X_{db}$ and sheet music $y_{k}$ from $Y_{db}$. The joint probability $p_{\theta}(J(x, y)|z)$ of audio and sheet music is a mean function, which is defined as follows.
\begin{eqnarray}
  p_{\theta}(J(x, y)|z) =\frac{exp(-||H(z)-J(x, y)||^{2})}{\sum_{J} exp(-||H(z)-J(x, y)||^{2})}\\
  J(x, y) = 0.5*(F(x)+G(y))
\end{eqnarray}
Where the final $p_{\theta}(J(x, y)|z)$ decides whether the possibility of an audio-sheet music pair is relevant or not.
\subsection{Discriminative Model}
We apply KNN method to exploit underlying manifold structure for the CCA embeddings of audio-sheet music pairs, we select the top five most close items to establish new pairs as ground truth. The input of the discriminative model is the generated audio and sheet music pair, and the manifold structure based ground truth. 



The target of discriminative model D is to distinguish the input audio-sheet music pair is from ground truth or generated. Once the discriminative model receives the two kinds of input pairs, the model will receive a relevance score for each pair (query and instance $i$) as the judgement score. The relevance score of $\varphi(p_{i}, q)$ is calculated by the following formulation:
\begin{eqnarray}
\Resize{8cm}{\varphi(p_{i}^{G},q)=max(0,\alpha+||\Theta(q)-\Theta(p_{i}^{T})||^{2}-||\Theta(q)-\Theta(p_{i}^{G})||^{2})}\\
 \Theta(x) = tanh(w_{i}x + b_{i})
\end{eqnarray}
where $q$ is the query, $p_{i}^{G}$ is the generated instance $i$ , $p_{i}^{T}$ is the ground truth. $\alpha$ is the margin parameter and it is set as 1 in our experiment. $w_{i}$ is the weight and $b_{i}$ is the bias.

The discriminative model applies the relevant score to calculate the predicted probability of a audio-sheet music pair $(x, y)$ by a sigmoid function.
\begin{eqnarray}
 D(p_{i}|q) = sigmoid(\varphi(p_{i}^{G}, q)) = \frac{exp(\varphi(p_{i}, q)}{1+exp(\varphi(p_{i}, q)}
\end{eqnarray}

\subsection{Adversarial Learning}
Once the concepts of the G model and D
model are accomplished, they are trained together by applying a minimax game. Inspired by the GAN~\cite{NIPS2014_5423}, this adversarial process can be defined as follows.
\begin{eqnarray}
\Resize{8cm}{V(G,D) =min_{\theta} max_{\phi} \sum_{i=1}^{n} (E_{x\thicksim{p_{true}}(x^{T}|q_{i})}[log(D(x^{T})|q_{i})] }\\
+E_{x\thicksim{p_{\theta}}(x^{G}|q_{i})[log(1-D(x^{G})|q_{i})])}
\end{eqnarray}

Fig.~\ref{fig:model} shows the architecture of our algorithm.  
We apply three pre-trained models to extract high-level semantic features. Vggish model has the same setting as~\cite{zengdeep} for audio, VGG-like mode follows~\cite{dorfer2017learning} for sheet music, and skip-gram model follows~\cite{yuHCMTJ20} for lyrics.  The input of G model is the global average of extracted features. Learning aligned representation is to take lyrics as Teacher model to teach the audio and sheet music model to learn discriminative representation, then transfer the lyrics model into audio and sheet music by adversarial learning. The goal of D model is to distinguish whether the input pair is from generated or ground truth by computing the relevance score for judgement result of each pair.
\section{EXPERIMENTS}
\subsection{Dataset collection}
The available musical dataset with three modalities is limited, we collect our musical dataset followed by the work~\cite{yuHCMTJ20} according to two rules as follows: 

1) ensure that each syllable-note paired sample contains 20 notes, it keeps the former first 20 notes as a sample or first 40 notes as two samples.
2) remove the samples if the silence period between two connected notes is longer than 4 seconds.

Since audio and sheet music can be generated from music notes, it can satisfy our objective of musical data with three modalities establishment. We extend music note to music audio and sheet music, synthesizing audio with TiMidity++\footnote{http://timidity.sourceforge.net/.}, creating sheet music by Lilypond\footnote{http://lilypond.org/}. Some samples are shown in the Fig.~\ref{fig:samples}.

Finally, we apply Recall@K (K=1, 5, 10), Median Rank and Median Rank to evaluate our proposed architecture.

\subsection{Implementation Details}
Our model is implemented by TensorFlow. Audio feature is extracted by Vggish model, applying ASMCMR~\cite{dorfer2017learning} model to extract sheet music feature, and using skip-gram model extracted word-level and syllable-level features for lyrics. The dimension of feature in the common space is set as 128. Moreover, we train our DARLearning model in a mini-batch with batch size 64 for both generative and discriminative model, all the fully connection layers of audio and sheet music in G model and D model share the same structure but learn its own weights and bias.

\subsection{Results}
\begin{table*}
\begin{center}
  \caption{Cross-modal Retrieval Results on MTM Musical Dataset with R@1 and MedR metric.}
  \label{tab:two_metrics}
  \begin{tabular}{c|cccccccccccl}
    \hline
    Method &\multicolumn{2}{|c}{audio} &\multicolumn{2}{|c}{lyrics} &\multicolumn{2}{|c}{audio} &\multicolumn{2}{|c}{sheet music} &\multicolumn{2}{|c}{lyrics} &\multicolumn{2}{|c}{sheet music}\\
    &\multicolumn{2}{|c}{$\downarrow$} &\multicolumn{2}{|c}{$\downarrow$} &\multicolumn{2}{|c}{$\downarrow$} &\multicolumn{2}{|c}{$\downarrow$} &\multicolumn{2}{|c}{$\downarrow$} &\multicolumn{2}{|c}{$\downarrow$} \\
    &\multicolumn{2}{|c}{lyrics} &\multicolumn{2}{|c}{audio} &\multicolumn{2}{|c}{sheet music} &\multicolumn{2}{|c}{audio} &\multicolumn{2}{|c}{sheet music} &\multicolumn{2}{|c}{lyrics} \\ \cline{2-13} &R@1  &MedR &R@1 &MedR &R@1  &MedR &R@1  &MedR &R@1  &MedR &R@1  &MedR\\
    \hline
    Baseline &3.69 &838.0 &2.97 &891.0 &24.46 &562.0 &25.13 &502.5 &3.18 &963.0 &3.05 &982.0 \\
    Baseline-GAN   &3.92 &804.2 &3.18 &858.0 &26.01 &529.0 &25.74 &472.2 &3.21 &940.0 &3.36 &925.8\\
    Our model   &5.02 &715.6 &4.14 &716.0 &30.06 &352.4 &33.02 &330.8 &8.36&572.5 &9.95 &576.0\\
  \hline
\end{tabular}
\end{center}
\end{table*}

\begin{table}
\begin{center}
  \caption{Cross-modal Retrieval Results on MTM Musical Dataset.}
  \label{tab:cross-modal}
  \begin{tabular}{c|ccccl}
    \hline
    \multicolumn{6}{c}{audio-to-lyrics retrieval}\\\hline
    Methods &R@1 &R@5 &R@10 &MedR &MeanR\\
    \hline
    RANDOM~\cite{AytarVT17} &2.77 &5.53 &7.61 &7312.0 &7257.19  \\
    Our model &5.02 &5.51 &6.34 &715.6 &808.02 \\
    \hline
    \multicolumn{6}{c}{lyrics-to-audio retrieval}\\
    \hline
    RANDOM &2.69 &5.45 &7.59 &7316.5 &7257.31  \\
    Our model &4.14 &4.56 &5.21 &716.0 &797.00 \\
  \hline
  \multicolumn{6}{c}{sheet music-to-lyrics retrieval}\\ 
    \hline
    RANDOM &2.74 &5.48 &7.53 &7311.8 &7257.26 \\
    Our model &8.36 &14.24 &16.58 &572.5 &765.30\\
    \hline
    \multicolumn{6}{c}{lyrics-to-sheet music retrieval}\\
    \hline
    RANDOM &2.72 &5.51 &7.66 &7313.2 &7257.43  \\
    Our model &9.95 &14.26 &17.02 &576.0&767.82\\
  \hline
  \multicolumn{6}{c}{audio-to-sheet music retrieval}\\ 
    \hline
    RANDOM &2.84 &5.57 &7.50 &7310.0 &7257.16  \\
    Our model &30.06 &33.98 &35.02 &352.4&600.34\\
    \hline
    \multicolumn{6}{c}{sheet music-to-audio retrieval}\\
    \hline
    RANDOM &2.63 &5.49 &7.48 &7310.0 &7257.37  \\
    Our model &33.02 &34.12 &35.88 &330.8 &584.42\\
  \hline
\end{tabular}
\end{center}
\end{table}
\textbf{Baseline} is only discriminative model without generative model and adversarial learning. The Baseline model is trained by triplet loss. 

\textbf{Baseline-GAN} is expanded Baseline with adversarial training. The input of D model is the pre-trained model extracted features.

Our model projects three modalities into a common space to support the representation can be compared with each other. Some initial results show in the tables. In table~\ref{tab:two_metrics}, it verifies the effective of our proposed model for transferring learning, which proves our hypothesis is acceptable. In detail, our model can transfer the relationship of two modalities to another one modality. In table~\ref{tab:cross-modal}, the result of three groups of cross-modal retrievals (audio-lyrics, sheet music-lyrics, and audio-sheet music) show the feasibility of further improvement of our proposed model.
\subsection{Further Analysis}
The initial experimental results suggest that our model is viability to learn alignment representations for audio, sheet music and lyrics for cross-modal retrieval task. Instead of learning representations of two variable sets, our model learns only one shared subspace across three modalities. The learned representations can keep the modality-variant and the paired data should have similar representations. 

We expect that our model can surpass CCA model in each couple of cross-modal retrieval in the future. Currently, the shortages of our modal are as follows: i) the loss in G model is not good enough to generate new representation of each modality. Especially, the mean function as the joint probability for the generative probability may weaken the relationship between audio-sheet music by only considering the relevant positions of audio and sheet music. ii) some weights of the fully connection is close to zero. In the future, we would like to use new method to normalize the input features. Overall, it requires us to enhance the relationship between the audio-sheet music during transfer learning with some advanced joint probability, such as considering the local positions of audio and sheet music.
\section{CONCLUSION}
Modality-invariant and discriminative representations empower multimodal intelligence to manipulate unrestricted and real world environment. Learning aligned representation is critical for the next generation of multimodal intelligence to learn each cross-modal data on multimodal content. Learning aligned representation between two data modalities has reached outstanding achievement. In this paper, we introduce a representation learning model on three data modalities. The experimental results show the feasibility for align representation learning across three different music modalities. Even though there are no audio-lyrics and lyrics-sheet music pairs for our model training, the results demonstrate the alignment can be learned by modalities-level transfer learning.

An open issue for future research is to develop a new generative model which can enhance the relationship of audio-sheet music pairs.

\section*{Acknowledgment}
This work was supported by JSPS Grant-in-Aid for Scientific Research (C) under Grant No. 19K11987.

\bibliographystyle{IEEEtran}
\bibliography{IEEEexample}

\end{document}